\documentclass[fleqn, usenatbib, useAMS]{aa}
\usepackage{graphicx,amssymb,bm}
\usepackage{times,url}
\usepackage{subfig,xfrac}
\usepackage{float, Abbrev}
\usepackage[fleqn]{amsmath}  
\usepackage{color}
\usepackage[none]{hyphenat}

\usepackage{placeins}

\newcommand{\be}{\begin{equation}}
\newcommand{\ee}{\end{equation}}
\newcommand{\ba}{\begin{eqnarray}}
\newcommand{\ea}{\end{eqnarray}}

\newcommand{\sidm}{\sigma_{\rm DM}/m}
\newcommand{\cpg}{cm$^2$/g}

\def\simlt{\lower.5ex\hbox{$\; \buildrel < \over \sim \;$}}

\newcommand{\fig}{\begin{figure} \begin{center}}
\newcommand{\efig}{\end{center}\end{figure} }
\newcommand{\figs}{\begin{figure*}\begin{minipage}{180mm} \begin{center}}
\newcommand{\efigs}{\end{center}\end{minipage}\end{figure*} }
\def\simgt{\lower.5ex\hbox{$\; \buildrel > \over \sim \;$}}

\usepackage{graphicx} 

\begin{document}
\titlerunning{Super-sampled SIDM}

\title{DARKSKIES: A suite of super-sampled zoom-in simulations of galaxy clusters with self-interacting dark matter}

\author{David Harvey \inst{1}\thanks{david.harvey@epfl.ch}
          \and  Yves Revaz\inst{1} 
          \and Matthieu Schaller\inst{2, 3} 
          \and Aurel Schneider \inst{4}
          \and Ethan Tregidga \inst{1}
          \and Felix Vecchi \inst{1}
          }

\institute{
Laboratoire d’Astrophysique, EPFL, Observatoire de Sauverny, 1290 Versoix, Switzerland
\and Lorentz Institute for Theoretical Physics, Leiden University, PO Box 9506, NL-2300 RA Leiden, The Netherlands
\and Leiden Observatory, Leiden University, PO Box 9513, NL-2300 RA Leiden, The Netherlands
\and Institute for Computational Science, University of Zurich, Winterthurerstrasse 190, 8057 Zurich, Switzerland
}

\label{firstpage}

\abstract{
We present the ``DARKSKIES'' suite of one hundred, zoom-in hydrodynamic simulations of massive ($M_{200}>5\times10^{14}M_\odot)$ galaxy clusters with self-interacting dark matter (SIDM). We super-sample the simulations such that $m_{\rm DM}/m_{\rm gas}\sim0.1$, enabling us to simulate a dark matter particle mass of $m=0.68\times10^{8}M_\odot$ an order of magnitude faster, whilst exploring SIDM in the core of clusters at extremely high resolution. We calibrate the baryonic feedback to produce observationally consistent and realistic galaxy clusters across all simulations and simulate five models of velocity-independent SIDM targeting the expected sensitivity of future telescopes - $\sidm=0.,0.01,0.05,0.1,0.2$\,\cpg. We find the density profiles exhibit the characteristic core even in the smallest of cross-sections, with cores developing only at late times ($z<0.5$). We investigate the dynamics of the brightest cluster galaxy inside the dark matter halo and find in SIDM cosmologies there exists a so-called wobbling not observed in collisionless dark matter. We find this wobble is driven by  accreting mass on to a cored density profile with the signal peaking at $z=0.25$ and dropping thereafter. This finding is further supported by the existence of an anti-correlation between the offset between the BCG and the dark matter halo and its relative velocity in SIDM only, a hallmark of harmonic oscillation.}

\maketitle
\section{Introduction}

Evidence for the existence of a dark, gravitating mass that dominates the matter component of our Universe has been building for almost a century \citep{zwicky33,rotationcurves,separation,planckParsFinal}. Despite its abundance in our Universe, no detection is yet confirmed, with the simplest model of a cold and collisionless massive particle that interacts only via gravity still consistent with observations \citep[hereafter referred to as "CDM",][]{DEScosmology,kids450}. 

In a bid to test theories of dark matter, computational models containing many billion point particles were built. In this way, it is possible to see whether changes in the standard dark matter paradigm lead to plausible Universes.  For example, early {\it n}-body simulations of pure gravity showed that dark matter must be non-relativistic, with so-called ``hot dark matter'' being quickly ruled out \citep{EvolutionLSS,HDM} as it produced too few structures.  In fact, the cold and collisionless dark matter model has been incredibly successful at predicting the large-scale distribution of observed galaxies \citep{lss_review,lss_review_2}.

As observations improved so did our ability to simulate smaller scales, pushing {\it n}-body simulations to reproduce the inner properties of galaxies and galaxy clusters. As a result, discrepancies in the cold and collisionless dark matter model began to appear. For example, dark matter only simulations predicted the existence of many small structures in the halo of galaxies, coined the ``missing-satellite'' problem \citep{satelliteprob,klypin1999,moore1999}. Moreover, it appeared that dwarf galaxies harboured flat density profiles in the central parts of the halo, whereas simulations of dark matter predicted cuspy, dense central regions known as the `core-cusp' problem \citep{corecusp,corecusp1,corecusp2}. These problems were quickly followed by other discrepancies between observations and simulations including the ``too-big-to-fail'' problem \citep{toobigtofail}. For a long time these were used as a way to motivate extensions to the CDM paradigm. For example, self-interactions on the dark sector can lead to a dampening of the central region and the production of a dark matter core \citep[]{SIDMSim}. However, it was clear that more sophisticated simulations that included star formation, cooling and other ``baryonic processes''\footnote{A term that includes all known processes associated with the galaxy formation.} were required. By including these additional gas physics it was possible to reconcile many of the observed tensions  \citep[e.g.][]{solve_missing_satellite,baryonsolution}.

The accuracy of simulations to model the intricate processes of galaxy formation has now advanced to the point where we can simulate large-scale boxes of the Universe and produce realistic objects over many orders of magnitude \citep{illustrisTNG,flamingo,magneticum}. These massive simulations are able to simulate a host of different physics including Active Galactic Nuclei (AGN), neutrinos, and magnetic fields leading to accurate predictions of observables such as the stellar luminosity function and the gas mass fractions in groups and clusters.

\begin{figure*}
\begin{center}
\includegraphics[width=\textwidth]{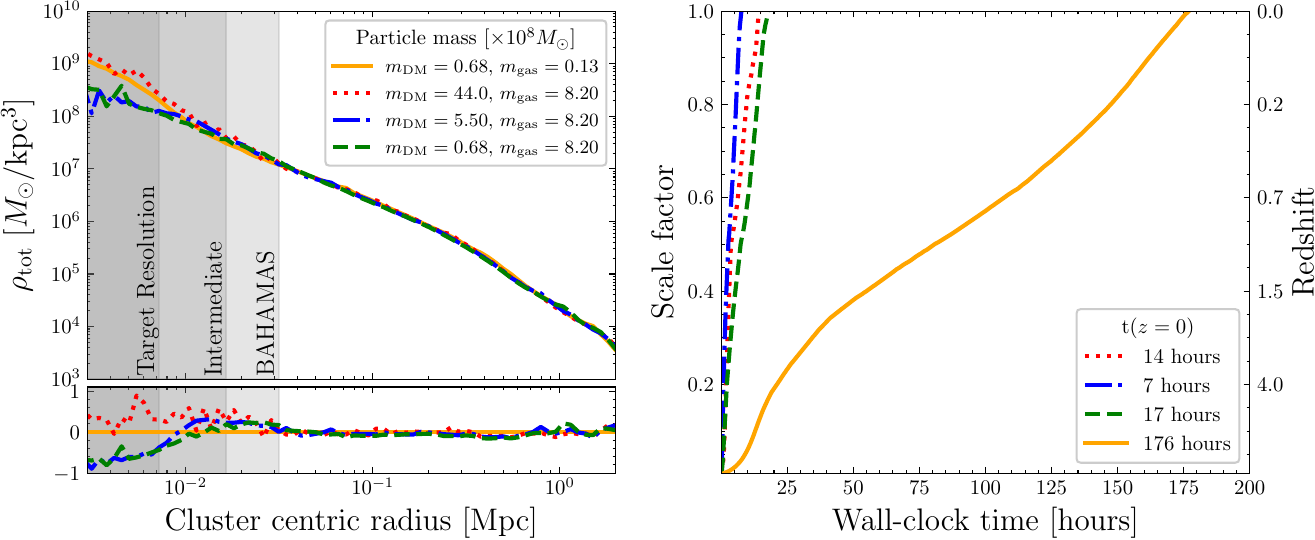}
\caption{ The validation check of our super-sampled initial conditions. By stitching multiple runs of {\tt MUSIC} we create non-standard combinations of dark matter and gas mass zoom-in simulations. We run five zoom-in simulations using the same initial conditions, increasing the dark matter mass resolution. We simulate each to a redshift of $z=1$ and show the total matter density profile here in the left hand panel with the relevant convergence radii in the grey regions (c.f. \eqref{eqn:convergence_rad}). We find consistent density profiles over the range of mass resolutions. The right hand panel shows the corresponding wall-clock time to simulate each halo to the same redshift. The colour and line-style match those in the left hand panel, we also show the final wall-clock time to simulate to $z=1$ in the legend.}
\label{fig:super_sampled}
\end{center}
\end{figure*}

Despite the progress in large-scale simulations, two ``small-scale problems'' arguably persist. The first is the diversity of rotation curves in galaxies. Current models struggle to predict the variance in the observed rotation curves of galaxies \citep{galaxy_diversity}. The second is the apparent overly-concentrated halos - whether it is manifested as an anti-correlation of the central dark matter density of  dwarf galaxy and their peri-centre distance \citep{Kaplinghat2019,overdense_halo_a} (although argued that this anti-correlation is not as strong as previous claimed \citep{against_corr}). Solutions to these discrepancies could possibly lie in the observations or in the simulated model of CDM and baryons. However, it is possible to relax the assumption that dark matter is collisionless and introduce self-interacting dark matter (SIDM) as a potential solution \citep{SIDM_diversity,SIDM_diversity_A}. Although a low cross-section ($\sim10$ \cpg) cannot alter dwarf galaxies in an observable fashion \citep{Harvey_dwarf}, invoking a large self-interaction in the dark sector, increases the speed at which the dark matter halo is tidally stripped, which can lead to a process called ``gravo-thermal collapse'' \citep{gravothermal_collapse_a, gravothermal_collapse_b, gravothermal_collapse}. In this situation the cooling of the core caused by self-interactions can move energy out of the central region, drawing in more particles and leading to further interactions and cooling. This results in a run-away effect whereby you have an overly dense region, more than that in CDM. Thus SIDM can produce both cores {\it and} cusps. However, some models suggest that even with gravo-thermal collapse SIDM is unable to recover the observed over-concentrated halos \citep{dark_dot_sidm}.

Gravo-thermal collapse requires a relatively high cross-section in dwarf galaxies (at the scale of $\sigma_{DM}/m \sim 100$ cm$^2$/g). Whereas observations of galaxy clusters have already ruled out this out with limits of $\sigma_{DM}/m < 0.5$ cm$^2$/g \citep{bulletcluster,impactpars,Harvey15,Sagunski}. Therefore, in order to reconcile these two limits, dark matter must be velocity-dependent, such that particles moving in clusters of galaxies (with masses greater than $M/M_\odot>10^{14}$) have small cross-sections, and particles in dwarf galaxies ($M/M_\odot<10^{11}$) have large cross-sections.

Despite the tentative detections of SIDM in dwarf galaxies, both the limited number of objects and the number of simulations to interpret the data remains the largest obstacle to progress.  Galaxy clusters on the other hand are numerous and benefit from their ``gravitational lensing effect''. The  mass of a galaxy cluster heavily deforms the space-time around it, distorting the observed images of distant galaxies in to correlated arc like structures \citep[see e.g.][for a review]{BS01}. In rare cases clusters can lead to multiple images of the same galaxy, known as strong gravitational lensing. Both weak and strong gravitational lensing probe the structure of all mass along the line-of-sight, with no assumption of its dynamical state.

Gravitational lensing has been the root of a host of constraints on the self-interaction, for example the first robust constraints\footnote{Constraints using cluster sphericity had previously been claimed, however subsequent simulations showed that observations were consistent with simulations \citep{SIDMTest,SIDMCore}} were from weak gravitational lensing of the bullet cluster \citep{separation}. Following this, studies of merging clusters \citep[e.g][]{A2744,minibullet,cannibal,Harvey15} with the tightest constraints from \citet{Sagunski} who used a sample of clusters and groups to constrain it to $\sidm<0.35$\,\cpg~, however, did so without simulations of SIDM. There have been some observations of potential signals of SIDM in clusters with measurements of the offset between dark matter and baryons in relaxed galaxy clusters - a smoking gun for dark matter self-interactions where offsets cannot exist in CDM \citep{LCDM_offsets}. \citet{Harvey_BCG} used 10 strong lensing galaxy clusters to measure a wobbling amplitude of $A_w=11.82^{+7.3}_{-3.0}$kpc. Similarly both  \citet{Harvey_BCG} and \citet{wobbles_tng} found that offsets in clusters between the dark matter halo and BCG are consistent with the softening length of the simulation. Indeed \citet{Harvey_BCG} attempted to model and deconvolve the impact of softening in order to achieve sub-smoothing length precision. Ultimately they found that the offsets observed in \citet{Harvey_BCG} were consistent with zero. However, de-convolving resulted in uncertain predictions and hence without high-resolution SIDM simulations it would be impossible to place constraints. This is particularly important as the number of observed clusters will soon rise dramatically with the first observations of the \textit{Euclid} satellite.

Numerical simulations of SIDM have been responsible for the revolution of this area of research. Thought to be ruled out by measurements of halo sphericity \citep{SIDMTest}, the first large-scale ($L>100$Mpc/h) cosmological simulations of SIDM found that cross-sections of $\sigma_{\rm CM}/m\sim1$cm$^2$/g were consistent with data \citep{SIDMSim,SIDMSimA}. Following this numerical studies exploring the observable manifestations of SIDM were exclusively dark matter only \citep[e.g][]{SIDMSimB}. However, these were limited by their exclusion of baryonic physics, which has a dramatic impact in the core of galaxies and clusters of galaxies. This was changed by the first large-scale, cosmological simulations of SIDM with full baryonic feedback \citep{RobertsonBAHAMAS}. Based on the BAHAMAS simulations \citep{BAHAMAS,BAHAMASB}, BAHAMAS-SIDM simulations were a suite of cosmological simulations based on the EAGLE model \citep{eagle} of star formation, cooling and AGN feedback (which itself was based on the OWLs simulation \citep{OWLS}. These were calibrated to reproduce the correct gas fraction in groups of galaxies and consisted of a $400$\,Mpc/h box, with dark matter mass of $m_{\rm DM}=3.85\times10^{9}M_\odot$ and an initial gas mass of $m_{\rm gas}=7.66\times10^{8}M_\odot$ and a WMAP9 \citep{WMAP9} cosmology. These were run with an adapted {\tt Gadget-3} smooth particle hydro-dynamics (SPH) solver \citep{gadget2} and a gas smoothing length of $\epsilon=5.7$ kpc. At this scale, they were able to resolve the interactions of dark matter down to the $\sidm=0.1$ \cpg. However, even at these scales the imprint of the softening could be observed on the characteristic scale at which it was interacting.  More recent work have looked to run zoom in cluster simulations with SIDM. For example, the DIANOGO simulations  looked at six high-resolution zoom in simulations \citep{Dianoga} where they tested two different models of SIDM: rare (with high momentum exchanged) and frequent (but low momentum exchange) dark matter self-interactions. Simulations of lower mass halos have also been well explored. For example, the AIDA simulations were extremely high-resolution cosmological boxes of five different dark matter models including Warm Dark Matter and SIDM \citep{AIDA}.  Leveraging the Illustris-TNG galaxy formation model,  AIDA-TNG looks at how dark matter affects the build up of intermediate mass halos \citep{AIDA}. In similar fashion, Concerto suite of SIDM simulations looked to trace the evolution of SIDM in zoom-in simulations in again intermediate mass halos \citep{concerto} at extremely high resolution ($m_{\rm DM} = (10^3 - 10^5)M_\odot$). Finally, SIDM simulations at very small scales have been extensively explored as the dark matter dominated dwarf galaxies present an exciting environment. The first simulations of SIDM in dwarf galaxies were run by \cite{SIDM_FIRE} and \cite{Harvey_dwarf}. Following this, further tests with velocity dependent cross-section looked at gravo-thermal collapse on the dwarf galaxy scale \citep{correa2022}.

We build on this body of work and present the first suite of high-resolution dark matter zoom-in simulations of massive galaxy clusters with a super-sampled distribution of dark matter, with a full prescription of independently calibrated baryonic feedback. These simulations are part of the ``DARKSKIES'' project aimed at constraining the self-interaction cross-section of dark matter to within $\sidm<0.1$\,\cpg. In section \ref{sec:sims} we present the underlying framework and model of our simulations, including how we generate the initial conditions, how we calibrate the meta-parameters that govern the hydro-dynamics and which models of SIDM we choose to simulate. In section \ref{sec:results} we present the results of the simulations and finally in \ref{sec:conclusions} we present our conclusions. Unless otherwise stated all error bars show the $1\sigma$ (68\%) confidence limit in a value, we also assume a Planck cosmology throughout \citep{planckParsFinal}.

\section{Method}\label{sec:sims}

Zoom-in simulations whereby the mass resolution of the system in a particular region is enhanced is an efficient way to simulate individual objects of interest, at high-resolution without wasting computational time on the large-scale surrounding environment. In this section we outline the methodology used to generate the initial conditions and then carry out the simulations, specific details of how we carry out the simulations follow, however as an overview, we carry out the following procedure:
\begin{enumerate}
    \item We run an initial $400h^{-1}$Mpc large-scale dark matter only (DMO) cosmological box from which to select the clusters to re-simulate.
    \item We select 30 galaxy clusters range in mass from $10^{14}<M/M_{\rm \odot}<10^{15}$, and trace back their particle positions to the initial conditions. From this we cut out a ``Lagrangian region'' (corresponding to five virial radii, $5r_{\rm vir}$) in which we want to setup as a high resolution region.
    \item We then generate super-sampled initial conditions using {\tt Music} \citep{music} and simulate 30 collisionless dark matter galaxy clusters, varying the hydro-parameters in order to produce viable galaxy clusters.
    \item Having calibrated these parameters we then return to the origin DMO box and select the one hundred most massive clusters in the simulation and re-simulate over a variety of SIDM parameters.
\end{enumerate}

 We use the smooth particle hydrodynamics with task based parallelism code {\tt SWIFT} \citep{swift} (specifically {\tt v0.9}), to carry out our zoom-in simulations.
 For the hydro runs we adopt the EAGLE sub-grid model of star formation and feedback based on \citet{eagle}, which was itself based on the OverWhelmingly Large simulations projects \citep[OWLs]{OWLS}. We use the open-source SWIFT-EAGLE galaxy formation model as detailed in \citet{swift-eagle} and \citet{swift-eagle-2}. Specifically, we use the same radiative and cooling rates from \citet{ploeck_2020}, which themselves used {\tt cloudy} code responsible for radiative transfer \citep{cloudy}. We use the \citet{OWLSstarFormation} pressure law for the stochastic star formation, ensuring that the star formation rate follows the Kennicutt-Schmidt law. We defined a gas particle to form stars if the sub-grid temperature is $T<1000$K, or if its density of hydrogen particles is $n_H>10^{-3}$ and the temperature is $T<10^{4.5}$K. Stars are born from an Chabrier initial mass function \citep{ChabrierIMF} between $0.1$ and $100{\rm M_\odot}$ and supernova feedback is stochastic, with stars between $8-100{\rm M_\odot}$ exploding via core collapse \citep{OWLSagn}, with heat coupled to the surrounding gas via the \citet{supernova_gas} method. As stars explode they release metals in to the intra-stellar-medium, whereby we track 9 metals including Hydrogen, Helium, Carbon, Nitrogen, Oxygen, Neon,
Magnesium, Silicon and Iron.

\begin{figure}
\begin{center}
\includegraphics[width=0.45\textwidth]{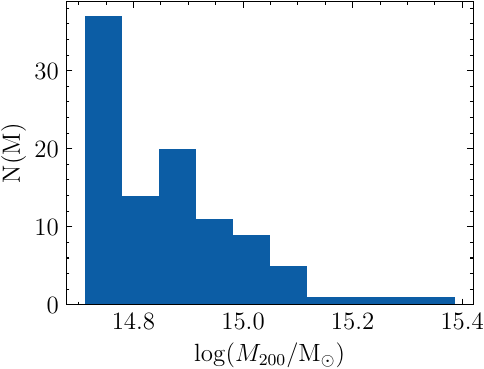}
\caption{ The distribution of total mass ($M_{200}$), of the 100 most massive galaxy clusters selected from the initial $400$\,Mpc$/h$ box to be re-simulated. }
\label{fig:mass_dist}
\end{center}
\end{figure}

\begin{figure*}
\begin{center}
\includegraphics[width=\textwidth]{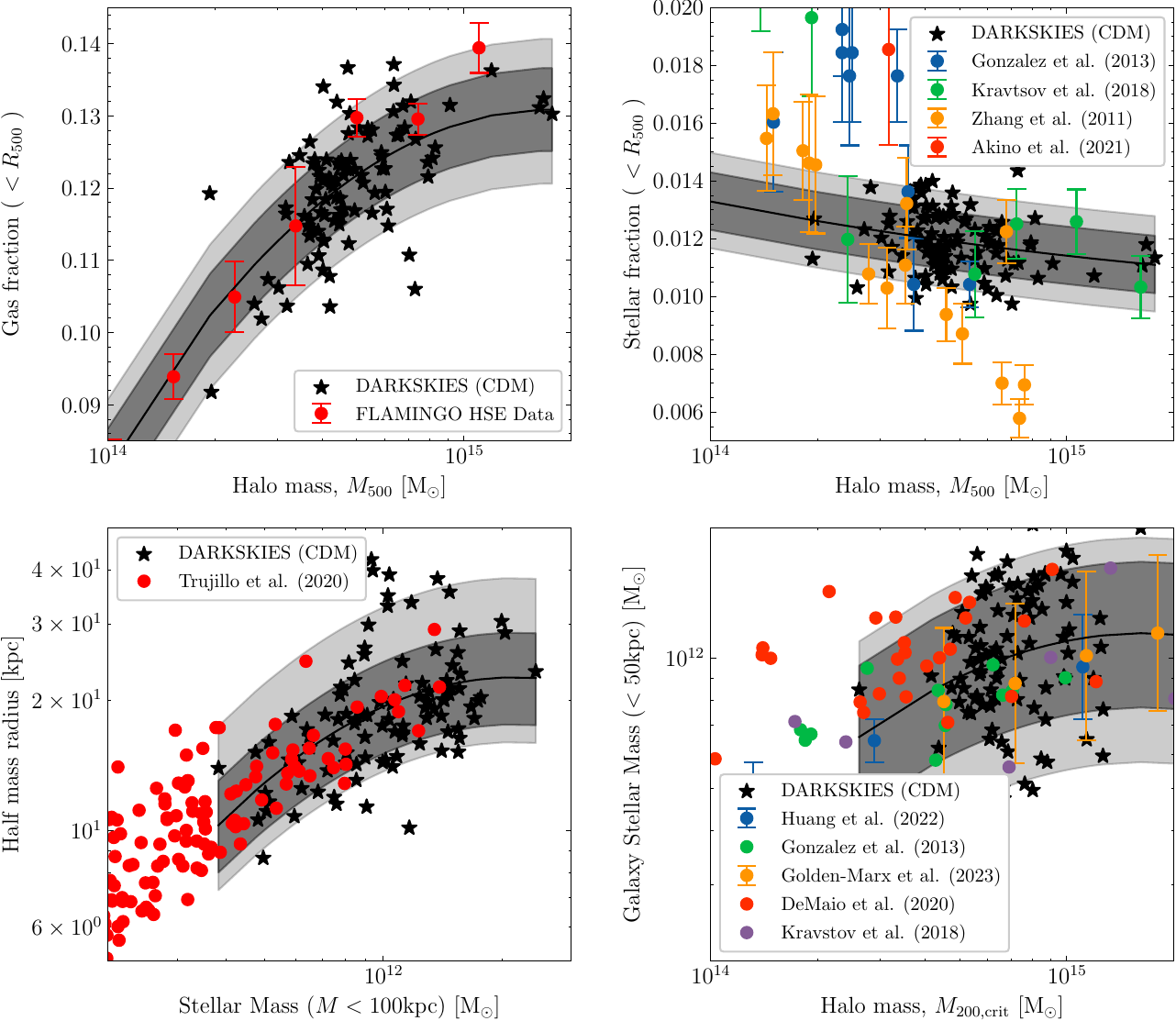}
\caption{ Final calibrated set of DARKSKIES CDM clusters. In each panel we show the zoom-in, super-sampled clusters with CDM in black stars. Specifically, {\it top left:} the gas fraction inside $r_{500}$ as a function of the halo mass $M_{500}$, {\it top right:} the stellar fraction inside $r_{500}$ as a function of the halo mass $M_{500}$, {\it bottom left:} the projected half mass radius of the central galaxy (considering all stars inside $100$kpc) as a function of projected stellar mass within $100$kpc, {\it bottom right:} The projected stellar mass within $100$kpc as a function of halo mass. The Flamingo HST Data corresponds to a collection of gas mass estimates collated by the Flamingo team \citep{gas_fraction,gas_fraction_a,gas_fraction_b,gas_fraction_c,gas_fraction_d,gas_fraction_e,gas_fraction_f}. The solid line in each case shows the best fit and the regions show the 68\% and 95\% around this fit.}
\label{fig:calibration}
\end{center}
\end{figure*}

Black holes are seeded at the centre of  friends-of-friends halos with a minimum mass of $M_{\rm FOF}=5\times10^{11}$ at a maximum redshift of $z=19$. Their formation and evolution follows the prescription in \citet{swift-eagle}. We assume a modified Bondi accretion rate that is limited to the Eddington rate as in \citep{OWLSagn2}, however as shown, since we do not resolve the multi-phase ISM, the accretion rate must be boosted by a factor of, $\alpha$:
\be
\alpha=
\begin{cases}
    1&  n_{\rm H}< 1\\
    \left(\frac{n_{\rm H}}{n^*_{\rm H}}\right)^{\beta_{\rm BH}}              &  n_{\rm H}\ge 1,
\end{cases}
\ee
 where $\beta_{\rm BH}$ is a parameter to be calibrated. Energy from the AGN is fed back in to the surrounding gas with some predefined efficiency, $e_f=0.15$. However, this energy is not injected at each time step, since this has been show to result in overcooling \citep{OWLSagn2}. We therefore build up energy in a reservoir until it reaches a threshold value, $\Delta T_{\rm AGN}$, at which time it is released into the surrounding gas particles. Both the size of this reservoir,  $\Delta T_{\rm AGN}$, and the number of gas particles coupled to this energy is a free parameter to be calibrated (see Section \ref{sec:calibration}).

 \subsection{SIDM Implementation}

We use the publicly available {\tt TANGO-SIDM} implementation SIDM scheme \citep{Correa2021,correa2022,correa2024}. {\tt TANGO-SIDM} is a probabilistic method that searches for the nearest dark matter neighbours and derives the probability of scattering between two particles, $a$ and $b$ based on their relative velocity and distance via:
\be
P_{ab}=m_b(\sigma_{\rm DM}/m)|\vec{v_a}-\vec{v_b}|g_{ab}(\delta\vec{r}_{ab})\Delta t,
\ee
where
\be
g_{ab}(\delta \vec{r}_{ab}) = N \int_0^{{\rm max}(h_a,h_b)} d^3\vec{r'}W(|\vec{r'}|,h_a)W(|\delta\vec{r_ab}+\vec{r'}|,h_b).
\ee
Here, N is the normalisation that ensures $\int_0^{{\rm max}(h_a,h_b)} d^3\vec{r'}g_{ab}(\delta \vec{r}_{ab})=1$, W is particle smoothing kernel with a width of $h_i$, $\sigma_{\rm DM}/m$ is the velocity independent cross-section, and $\Delta t$ the time-step. $h_a$ and $h_b$ are also the search radii within which  {\tt TANGO-SIDM} search for nearest neighbours to scatter with. Similar to smooth particle hydrodynamics, this radii is adaptive, depending on the local dark matter density, differing from many prescriptions (such as \citet{RobertsonBAHAMAS}), that used a fixed search radius. Having determine the probability of scattering, it compares to a random number to determine whether the scattering is successful, with any scatterings occurring elastically, isotropically, and conserving momentum. In order for these probabilities to remain valid we must avoid having multiple scatterings in a single time-step. Thus the time-step is reduced such that we expect only one scattering per time-step:
\be
\Delta t_i < \kappa \times \left[ \rho_a \langle \sigma_{\rm DM}/m \rangle (v_a)\sigma_{v,a}\right]^{-1},
\ee
where $\kappa=10^{-2}$, $\rho_a$ is the density of the dark matter particle, $a$, $ \langle \sigma_{\rm DM}/m \rangle$ is the average cross-section with respect to its neighbours (which in our case of a a velocity-independent cross-section is simply $\sigma_{\rm DM}/m$), and $\sigma_{v,a}$ is the velocity dispersion of particle $a$. This requirement can lead to dramatic increases in the computational time since $\Delta t$ can become much smaller than the typical time-step of the simulation, particularly in dense regions (or simulations with high cross-sections). However, in our case cross-sections are very low so we do not expect dramatically lower computational times.
For a complete derivation including associated equations please see \citet{correa2022} and the Appendices within.

\subsection{Initial conditions, Mass resolution and super-sampling}

We first run a dark matter only cosmological simulation in a large-box from which to select our clusters to re-simulate. How large this box is will be defined by the sample of clusters we wish to re-simulate - which in itself is a trade off between the size of the cluster and how long each cluster will take to re-simulate (with larger clusters with more particles requiring longer to simulate). In order to match the sample to the BAHAMAS cluster sample we simulate a box of $400~$Mpc/h with 256$^3$ particles and an initial redshift of $z=127$. We generate initial conditions for this box using the publicly available {\tt MUSIC} \citep{music} assuming a Planck cosmology \citep{planckParsFinal}. We then use the publicly available {\tt VELOCIraptor} code \citep{velociraptor} to find all halos within the field. For a given selected halo we find all the particles that are within a radius of $r<5r_{\rm vir}$ of the most bound particle and find their positions in the original initial conditions. We then trace a Lagrangian region around these particles and sample it at much higher resolution with {\tt MUSIC} to generate our zoom-in initial conditions. 

\begin{figure*}
\begin{center}
\includegraphics[width=\textwidth]{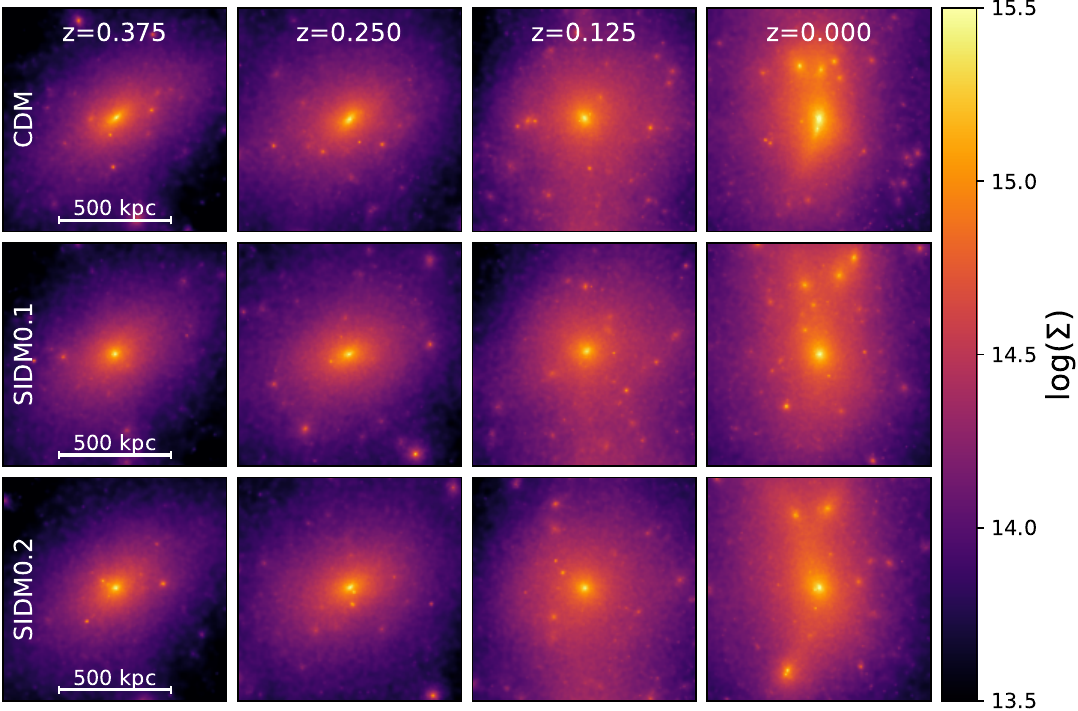}
\caption{ The total mass density map for the halo 20926 at different redshifts (columns), simulated in different dark matter models (rows). Each panel has a field-of-view of $1\times1$\,pMpc and has the same scaled colour bar and has been projected along the z-axis of the simulation box. }
\label{fig:sample_cluster}
\end{center}
\end{figure*}

In order to probe cross-sections at the scale $\sidm<0.1$ \cpg\, we require a dark matter mass resolution significantly greater than that of the BAHAMAS simulations. The Plummer-equivalent softening length, $\epsilon$, of a simulation roughly goes as $\epsilon=L/(25N)$ (where L is the box length, N is the number of particles i.e. a 25th of the inter-particle spacing), thus if we want to resolve scatterings on the scale of $\sim1h^{-1}$kpc, in a galaxy cluster with a target-resolution region size of around $5$\,Mpc we require a mass of the order $M\sim10^7M_\odot$. At this resolution, we would require a gas mass of $M_{\rm gas}\sim10^6M_\odot$, matching that of the \citet{ceagle} and \citet{CEAGLE_SIDM}, which would be prohibitively expensive, with individual simulations taking many months to run. Instead we choose to keep the initial gas mass the same as in the BAHAMAS simulations. Since it is the baryonic processes that contributes significantly to the core CPU time, this should significantly decrease the required simulation time, whilst retaining the required mass resolution to resolve small cross-sections. Of course one could reduce the softening length of the simulation in order to probe smaller scales in the simulation (for example \cite{millennium} used 1/42), however this would lead to greater CPU times and recalibration. Future work should understand the interplay between the softening length and the effect of SIDM.
\begin{table*}
    \centering
    \begin{tabular}{ccccccccc}\hline
        Identifier &$ m_{\rm dm}$  & $m_{\rm g}$ & $\sigma_{\rm DM}/m$ & $\Delta T_{\rm AGN}$  & $\beta_{\rm BH}$ & $n_{\rm AGN}$ & $\epsilon_{\rm prop}$ & \# Halos \\
         &  [$\times 10^8 $M$_\odot$] & [$\times 10^8 $M$_\odot$] & [cm$^2$/g] & [K]  &  &  & [pkpc ]& \\
        \hline
         CDM &  $0.68$ &  $8.2$ & $0$ & $10^8$ & $0.5$ & $20$  & $2.28$ & $100$  \\
         SIDM0.01&  $0.68$ &  $8.2$ & $0.01$ & $10^8$ & $0.5$ & $20$  & $2.28$ & $100$  \\
         SIDM0.05&  $0.68$ &  $8.2$ & $0.05$ & $10^8$ & $0.5$ & $20$  & $2.28$ & $100$  \\
          SIDM0.1&  $0.68$ &  $8.2$ & $0.1$ & $10^8$ & $0.5$ & $20$  & $2.28$ & $100$  \\
         SIDM0.2&  $0.68$ &  $8.2$ & $0.2$ & $10^8$ & $0.5$ & $20$  & $2.28$ & $100$  \\
         \hline
    \end{tabular}
    \caption{Overview of the key simulation parameters including (from left to right), the identifier used throughout this work, the mass of the dark matter particle, $m_{\rm dm}$, the initial gas particle mass, $m_{\rm gas}$, the self-interaction cross-section, the temperature reservoir, $\Delta T+{\rm AGN}$, the power index for the black hole boost factor, $\beta_{\rm BH}$, the number of gas particles the AGN energy release couples to, $n_{\rm AGN}$, the Plummer equivalent gravitational softening length at $z=0$, $\epsilon_{\rm prop}$ and the number of re-simulated, zoom in halos.}
    \label{tab:simulations}
\end{table*}
In order to generate initial conditions (ICs) of these type we run {\tt MUSIC} twice, the first at the required resolution for the gas mass and a second time for target resolution of the dark matter. We then  simply extract the gas particles (along with their positions, masses and velocities) from the low resolution run and concatenate them with the dark matter particles from the high resolution run to create a new set of initial conditions.  In order to validate this method we simulate a single halo, with a set on fiducial initial conditions and slowly increase the dark matter resolution and simulate down to a redshift of $z=0$. We assume an uncalibrated set of hydro-parameters in order to test the consistency of the initial conditions, with a $\log(\Delta T_{\rm AGN}) = 8.0$, and an $\beta_{\rm BH}=0.8$. We measure the total density profile of the halo and report the results in the left hand panel of Figure \ref{fig:super_sampled}. We also show the various convergence radii as defined by \citet{densityprofSIDM} and show how they correspond to different simulations. We begin by simulating a halo from a single {\tt MUSIC} run at the resolution of the BAHAMAS (red dotted line), and then decrease the dark matter mass such that they are roughly equal (blue dot-dashed), then at the target resolution of $m_{\rm gas}=10m_{\rm dm}$. We then simulate  (green curve) a full, super-high resolution in both gas and dark matter from a single set of initial conditions (solid yellow line) and find the density profiles match. In the right hand panel of Figure \ref{fig:super_sampled} we show the wall-clock time to reach the target redshift of $z=1$. We see that the improvement in time for base simulation a factor of two (colours match the legend in the right hand panel). However, if we were to include the respective gas mass resolution it would lead to a factor of 10 increase in time (yellow line, 86 hours). The legend shows the wall-clock time to $z=1$, with the colours matching those simulation parameters in the left hand panel.

Interestingly recent work has shown that a mismatch in particle mass (where often the dark matter mass is significantly larger than the stellar mass) in simulations can lead to numerical heating, artificially puffing up the halo \citep{spurious_heating,spurious_heating_b,spurious_heating_c}. Here we are in a different regime where the mean stellar mass is five times larger than the dark matter mass, potentially heating the dark matter and mimicking SIDM. We observe no such heating in the core of the clusters, potentially since the particle number of stars is so few that it wont have much impact. However, it is important to consider such an effect when super-sampling the initial conditions.

The final dark matter and initial gas mass resolution throughout the rest of this paper is $m_{\rm dm}=0.68\times10^8M_\odot$ and $m_{\rm gas}=8.20\times10^8M_\odot$ respectively (corresponding to level 10 and 12 in the {\tt MUSIC} IC generator). We adopted a Plummer-equivalent comoving softening length of $8.93$ckpc, and a maximum smoothing length of $2.28$pkpc for both the baryons and the dark matter. Confident that the initial conditions do not result in spurious numerical problems we now look to calibrate our simulations.

\begin{figure*}
\begin{center}
\includegraphics[width=\textwidth]{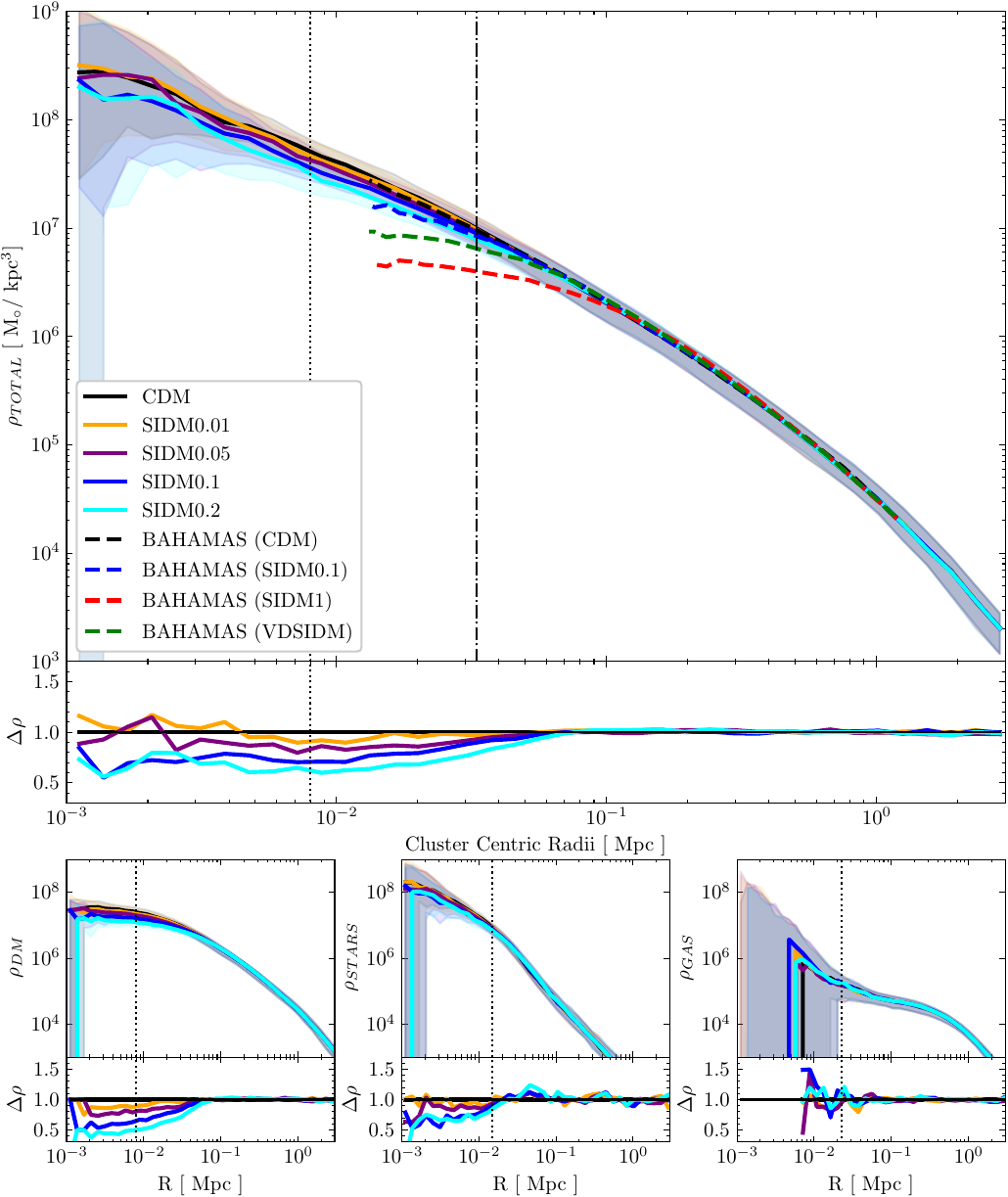}
\caption{ The total (top panel), dark matter (bottom left panel), stellar (bottom middle panel) and gas (bottom right panel) mass density profiles as a function of self-interaction cross-section. In each case we show, $\Delta\rho$ - the density relative to CDM and in the main panel we show the BAHAMAS SIDM simulations as dashed lines. In each case the shaded region shows the 16\% and 84\% of the simulations. }
\label{fig:density_profs}
\end{center}
\end{figure*}

\begin{figure}[!h],
\begin{center}
\includegraphics[width=0.5\textwidth]{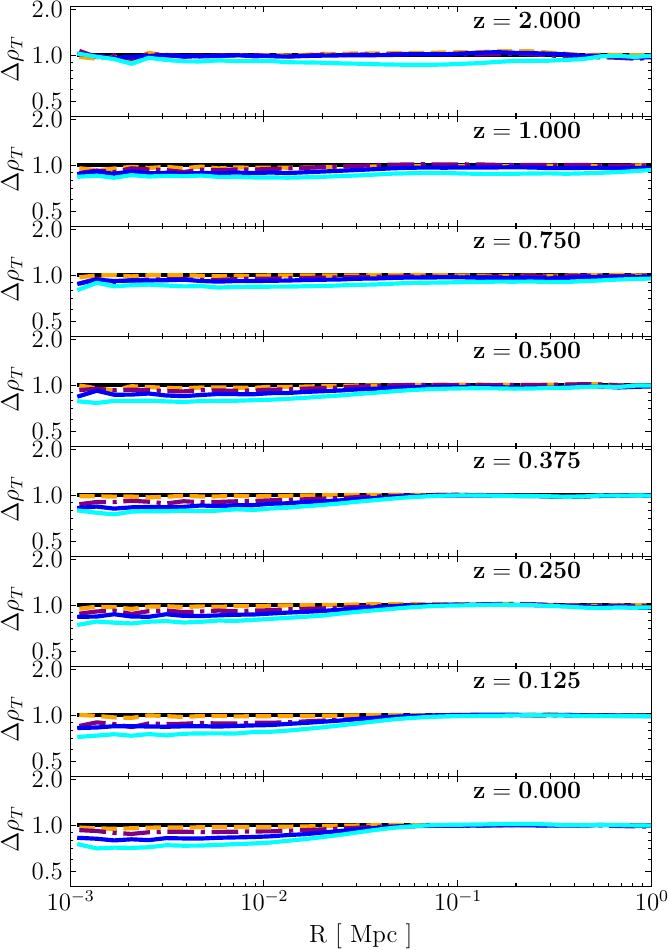}
\caption{ The redshift dependence of the total matter three dimensional density profile of the clusters in the DARKSKIES simulations relative to the CDM case. Each panel shows the density profile for each SIDM model at a given redshift slice. The colours are the same as in Figure \ref{fig:density_profs}.   }
\label{fig:redshift}
\end{center}
\end{figure}

\subsection{Calibration}\label{sec:calibration}

%Most cosmological simulations often fail to accurately produce the correct stellar fractions in very massive clusters of galaxies. This arises from the fact that often the simulations are calibrated on small boxes, that do not produce sufficient massive clusters to calibrate on. As a result many modern suites of simulations over-produce stars in the most massive clusters \cite[for example][]{flamingo,illustrisTNG,tngcluster}. 
The special setup of our simulations, with super-sampled dark matter requires us to recalibrate, since no hydro model has been tested in such an environment. Moreover, correctly calibrating the central brightest cluster galaxy is particularly important for SIDM since the interplay between the inner regions or clusters and the stars can dramatically alter the resulting density profile \citep{CEAGLE_SIDM}. For example, \citet{SIDM_BAHAMAS} found that the wobble BCG signal (a smoking gun of SIDM \cite{darkgiants}) was extremely sensitive to stellar content of the brightest cluster galaxy (BCG). This makes logical sense, since the self-interactions are trying to expel particles, yet, any large concentrations of stars will attempt to prevent this through gravitational drag and contraction effects coming from cooling of gas. As a result, we focus our calibration on the stellar fractions and size-mass relations of the BCGs, to ensure that we get accurate predictions of what SIDM does in the core of a cluster. 

%Using \citet{flamingo_calibrate}, 
We calibrate our zooms equally to four key observables, including the gas mass fraction \citep{gas_fraction}, the stellar mass fraction \citep{gold_stellar_mass,krav_stellar_frac, zhang_stellar_frac,akino_stellar_frac}, the BCG size-mass relation \citep{Trujillo_half_mass} and finally the stellar mass to halo mass function for BCGs \citep{gonzalez_stellar_frac,gold_stellar_mass,demaio_stellar_mass,krav_stellar_frac,huang_smhm}. These are all equally important since obtaining the correct stellar fraction, as mentioned, is a focus of this project. We do not calibrate to our final sample of 100 massive clusters as this would take too long - as such we choose a representative sample of 30 clusters equally spaced between $14<\log{M_{200,{\rm crit}}/{\rm M_\odot}}<15.5$ in mass. These clusters span a large mass range for two reasons - the first is to ensure that we have some handle of the mass dependence of our calibration decisions - and secondly smaller mass halos are significantly faster to simulate and hence more efficient when tuning individual parameters.
Although the simulations used to train the machine learning model in \citet{flamingo_calibrate} have a slightly different feedback model to the one used here, the general principals remain and hence we use this study to direct our calibration points, shifting the $\Delta T_{\rm AGN}$ and $\beta_{\rm BH}$ to recreate realistic haloes. We find for our sample the best fit hydro parameters are $\beta_{\rm BH}=0.5$, $\Delta T_{\rm AGN}=10^8$K. Finally, as in BAHAMAS, we choose to couple energy injections from AGN feedback to $20$ nearest neighbours. We test a variety of numbers and find that the properties are not sensitive to this choice.

Following the calibration of the hydro parameters, we simulate the 100 most massive clusters in the initial $400$\,Mpc$/h$ box. We show their mass distribution in Figure \ref{fig:mass_dist} and their final calibrated parameters in Figure \ref{fig:calibration}. We also quantify the fit and scatter of our simulated clusters to ensure they are consistent with observations.  The solid line in each case shows the best fit and the regions show the 68\% and 95\% around this fit. We find our cluster match extremely well the observations at this mass range. It is these clusters that will be re-simulated in a SIDM cosmology.
\begin{figure*}
\begin{center}
\includegraphics[width=\textwidth]{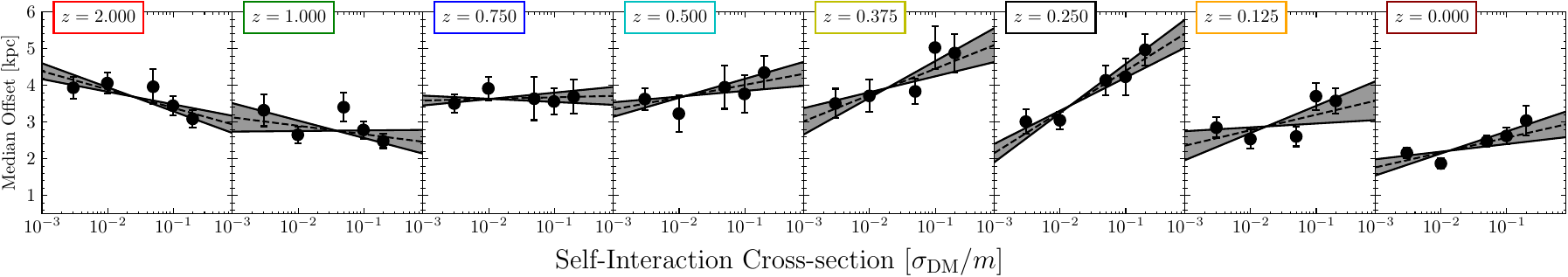}
\caption{ The redshift evolution of the three dimensional median offset between the dark matter halo and the stellar halo for each model of dark matter from $z=2$ (left) to $z=0$ (right).  We fit equation \eqref{eqn:wobble} to each snapshot and show the best fit with associated error in each panel.} 
\label{fig:wobble_z}
\end{center}
\end{figure*} 

\begin{figure}
\begin{center}
\includegraphics[width=0.48\textwidth]{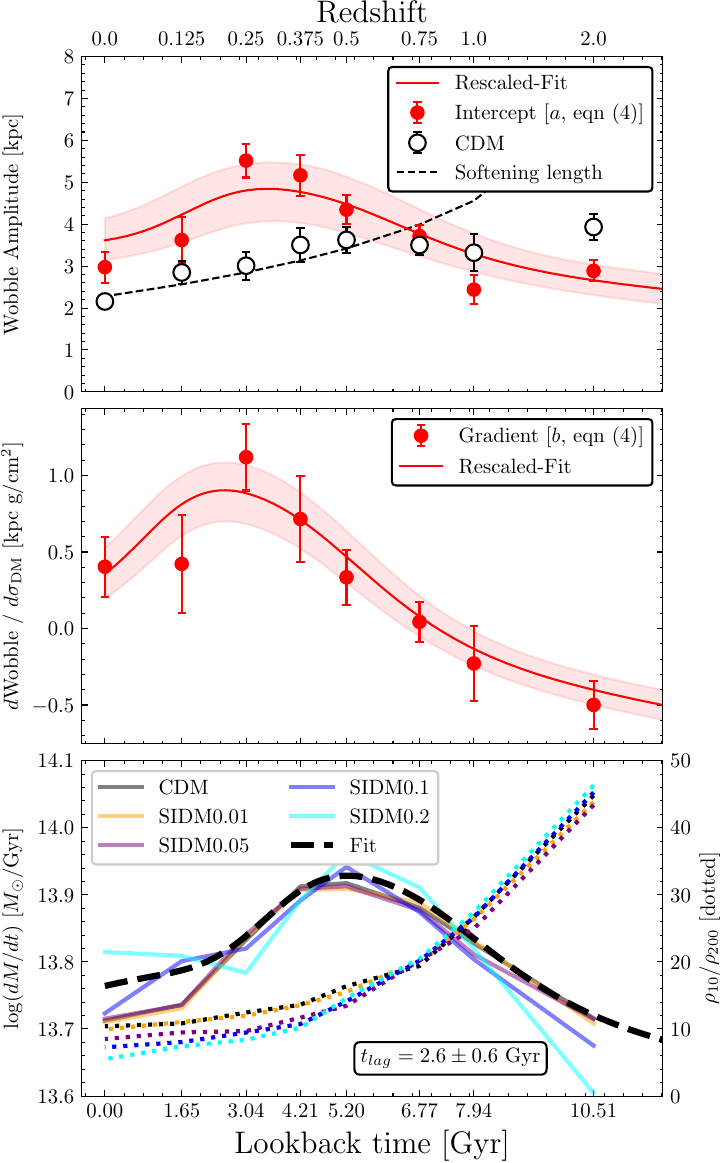}
\caption{ The time evolution of the fits to Figure \ref{fig:wobble_z} parametrised by equation \eqref{eqn:wobble}. The top panel shows the evolution of the intercept, $a$ plus the absolute values of CDM and the evolving smooth length of the simulation. The middle panel shows the evolution of the gradient $b$ as a function of look-back time (and respective redshift on the top axes). In both cases the error bars of individual points come from the estimated error from {\tt scipy.curve\_fit} and the shaded regions represent the $1-\sigma$ error in rescaled fit of the mass accretion curve. To determine these we Monte Carlo the fitted parameters around their uncertainties and draw random realisations of the fit and find their 16\% and 84\%. The bottom panel shows the mean mass accretion for each model of dark matter as a function of look-back time. The thick dashed line is the best fitted model to the mean mass accretion, which is subsequently fitted to the top two panels by stretching in x and y (see equation \eqref{eqn:refit}), with the best fit shown in red (and the shaded region the $1\sigma$ error. We also calculate the shift between the peak of the mass accretion rate and the peak of the wobble.  }
\label{fig:wobble_acc}
\end{center}
\end{figure} 

\subsection{SIDM Parameter space}

Optimally sampling the SIDM parameter space in order to provide a valuable dataset for a variety of scientific purposes is difficult. For example, when measuring classical parameters such as the offset between a central galaxy and the dark matter halo, a large sample of clusters simulated at the same cross-section is required in order to determine the overall distribution \citep{SIDM_BAHAMAS}.  However, providing samples at many different cross-sections is often more valuable since it can provide more variable data for a model to fit to \citep[e.g][]{camels}. As such we must balance the two contrasting requirements since satisfying both is computationally unfeasible. We also note that should SIDM be velocity-dependent then this introduces two new parameters \citep{CEAGLE_SIDM}. However, as it was found in \citet{RobertsonBAHAMAS}, halos at low redshift ($z<0.5$) at a given mass (and hence characteristic particle velocity), simulated at a velocity-independent cross-section is qualitatively the same as if it was simulated with a velocity-dependent cross-section that had a matching parametrisation of the normalisation, $\sigma_0$ and turn-over velocity, $w$, where the cross-section can be characterised, for a given dark matter model, as
\be
\sigma/m = \sigma_0/m \left( 1 + \left(\frac{v}{w}\right)^2\right)^{-1}.
\ee
\citep[see e.g][]{RobertsonBAHAMAS,correa2022}.
In particular, machine learning algorithms cannot tell them apart \citep{darkcnn}. This idea of an "effective cross-section" has been explored in \cite{effective_crosssection} (in particular equation (4.2)). 
 As such we choose to simulate only velocity-independent cross-sections, since these can approximately be recast into velocity-dependent realisations (since the direct mapping is non-trivial). Noting here that the caveat is that the sub-halos will not correspond to correct dark matter model, since in a velocity-dependent model we would expect changes in the internal structure where particles are moving slower relative to one another \citep{gravothermal_collapse}, hence, any observables are only valid for the main halo not sub-halos. Figure \ref{fig:par_space} shows the effective models that we are simulating here.

In order identify halos across models and look-back time we identify the main halo in each CDM, $z=0$ snapshot and trace back the particle ID's for each simulation. This enables us to carry out a like-for-like verification that the calibration is not dependent on the cross-section and that in all cases the properties of the clusters do not change with respect to the cross-section and hence does not require a re-calibration. Finally, we identify 5 cross-sections to simulate that are specifically chosen with the scientific goals of ``DARKSKIES''. In order to achieve precision of $\sidm<0.1$\,\cpg~ we require simulations that straddle this threshold. We therefore first select a cross-section that targets optimistic constraints expected from future telescopes such as Euclid using machine learning \citep{darkcnn} ($\sidm=0.01$\,\cpg), then a more conservative expectation at $\sidm=0.05$\,\cpg, the required limit of ``DARKSKIES'' of $\sidm=0.1$ \,\cpg\citep{Harvey_BCG, sirks2024} (and overlap with BAHAMAS-SIDM) and finally a larger cross-section at the current limit of what is observed $\sidm=0.2$\,\cpg, plus CDM ($\sidm=0$\,\cpg). We will refer to these models going forward as SIDM0.01, SIDM0.05, SIDM0.1, SIDM0.2 and CDM. A final overview of the simulations can be seen in Table \ref{tab:simulations}

    \begin{figure*}
    \begin{center}
    \includegraphics[width=\textwidth]{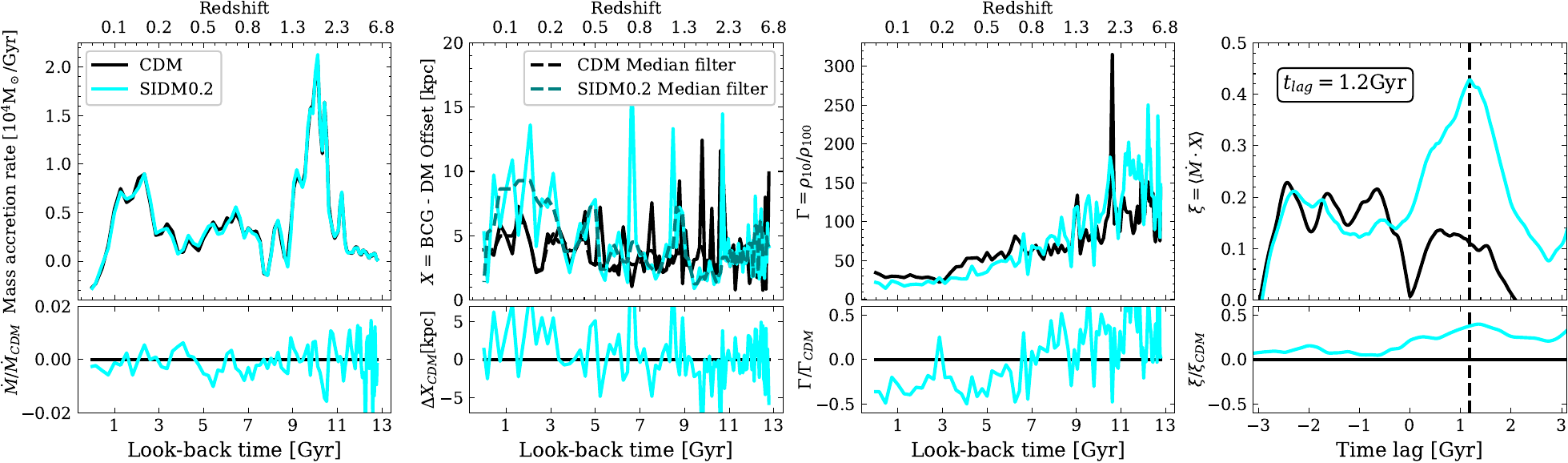}
    \caption{ High fidelity simulation of a single halo for two dark matter models. The left hand figure shows the mass accretion rate over cosmic time, with the bottom panel showing the accretion relative to CDM. The middle panel show the evolution of the offset between the dark matter and the brightest cluster galaxy, with a median filter (kernel size of 1 Gyr) shown in the dashed line. The oscillations / wobbles can be clearly seen at redshifts $z<0.5$. The bottom panel of the middle figure shows the difference between the offsets, showing how the SIDM0.2 grows at late times. Then the final, right hand figure shows the correlation between the mass accretion rate, $\dot{M}$ and the offset between the dark matter and the brightest cluster galaxy ($X$) as a function of the time lag between the signals. We see that for this halo the peak correlation occurs at 1.3 Gyr, consistent with the over distribution found in the bottom panel of Figure \ref{fig:wobble_acc}  }
    \label{fig:hig_fid}
    \end{center}
    \end{figure*} 
    
    \begin{figure*}
    \begin{center}
    \includegraphics[width=\textwidth]{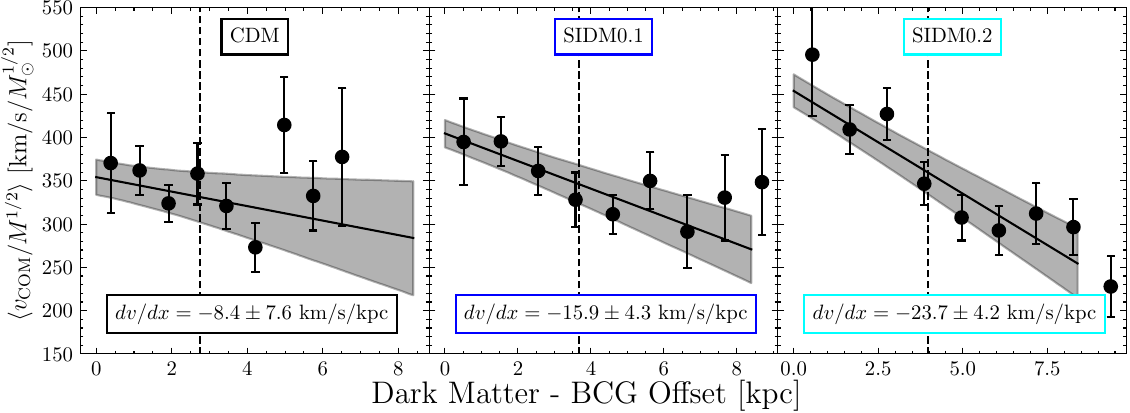}
    \caption{ The mean relative velocity of the BCG with respect to the main halo as a function of their three-dimensional spatial offset for three of our dark matter models for all halos $z<0.75$. We weight the velocities to the inverse square of the halo mass in order to stack. We show the wobble for each model with the vertical dash line. A key signature of harmonic oscillation is the anti-correlation between the velocity and the distance.}
    \label{fig:wobble_vel}
    \end{center}
    \end{figure*} 
    
\section{Results}\label{sec:results}

We now show the results from the simulations. We show an example cluster (ID=20926) in Figure \ref{fig:sample_cluster} for different redshifts (columns) and different dark matter models (rows). We see the stochastic nature of the simulations means that although the structure of the halo is the same the details vary from simulation to simulation. We next derive the three dimensional mass-density profiles for each set of clusters. In Figure \ref{fig:density_profs} we show the median profile for each run and each mass component. In the main panel we show the total density profile for each dark matter model simulated here in a solid line, with CDM in black, SIDM0.01 in orange, SIDM0.05 in purple, SIDM0.1 in blue and SIDM0.2 in cyan. We also show the convergence radii with the vertical dotted line \citep{densityprofSIDM}, where the convergence radii is defined as the smallest radii that satisfies
\be
0.33 \leq \frac{\sqrt{200}}{8}
\sqrt{\frac{4\pi\rho_{\rm crit}}{3m_{\rm DM}}}
\frac{\sqrt{N(r<r_{\rm conv})}}{\ln N(r<r_{\rm conv})}r_{\rm conv}^{2/3},
\label{eqn:convergence_rad}
\ee
where $\rho_{\rm crit}$ is the critical density at the redshift of the cluster, $m_{\rm DM}$ is the dark matter particle mass, and $N$ is the number of particles. The original convergence radii ensured that the mean density within this radius has converged to within 10\% of that of a higher resolution run \citep{convergence_test}. However, \citep{densityprofSIDM} relaxed this assumption since baryons naturally add increased scatter, thus they choose a radius whereby he mean density within this radius has converged to within 20\% of that of a higher resolution run. Theoretically there should be no reason why this criterium should not hold here, and thus we adopt the same criterium here despite having different particle mass.

For comparison, we show the  BAHAMAS+SIDM runs in dashed lines, with black showing CDM, blue showing SIDM0.1, red showing SIDM1 and velocity-dependent cross-section in green ( $\sigma_0/m=3.04$ \cpg, $w=560$km/s or $m_\chi=0.15$GeV, $m_\phi=0.28$keV and $\alpha=6.74\times 10^{-6}$), which at these mass scales corresponds to an effective cross-section of $\sim0.5$ \cpg. We show the convergence radii shown in the vertical dot-dashed line. Finally, we show in the bottom sub-panel the relative difference between each dark matter model and the DARKSKIES CDM run. We find that the BAHAMAS CDM runs agree extremely well with the CDM runs here. 

In the bottom three panels of Figure \ref{fig:density_profs} we show the three-dimensional density profiles of the individual three mass components, including (from left to right) the dark matter density profile, the stellar density profile and the gas profile, with the difference relative to CDM in the bottom, sub-panel. The vertical dotted line show the convergence radius for that mass component. We find that both the dark matter and stellar profiles are affected by the dark matter self-interactions, whereas the gas does not. However, because of the low-resolution of the gas it is difficult to make any firm conclusions from this plot.

We next look at the redshift dependence of the profile. Since scattering rates are linear in time, the signal should build up over cosmic time, with the largest differences at $z=0$. Figure \ref{fig:redshift} shows the density profile with respect to CDM at different redshifts and look-back times $t_{\rm LB}$. We find above a redshift of $z=0.5$ the signal for SIDM all but disappears, with the largest differences at lower redshift. However, there seems to be some suppression in the mass profile at intermediate radii for larger cross-sections ($10-100$kpc, SIDM0.2), suggesting a difference in mass build up of the cluster due to self-interactions. Interestingly, AIDA also find tentative evidence for a difference in the halo mass function for SIDM models at high-redshifts, suggesting that perhaps self-interactions are able to slow down the growth of structure.

Next we look at the offset between the dark matter halo and the BCG. Many studies have suggested that in a SIDM cosmology, any offset induced between the dark matter halo and BCG by a major merger at any point during the history of the cluster will persist, even long after the cluster has relaxed and virialised due to the flat potential in the core \citep{darkgiants,Harvey_BCG,SIDM_BAHAMAS}. This offset is a feature of SIDM and not CDM. Since we cannot observe this "wobbling" in real time, we can measure the one-point distribution of offsets between dark matter and the BCG and measure the median offset. The work in \citet{Harvey_BCG} identified the observable cosmological signal of the Wobble in two-dimensional images of the cluster and how it scaled with the cross-section. However, it wasn't clear whether this scaling was physically driven or as a result of increased error in a halo due to large dark matter cores. Here we want to identify very clearly, first the signal and its redshift dependence, and what processes are driving this.

To this end we first have to carefully identify the centre of mass of each cluster. This is not sufficiently precise in the {\tt Velociraptor} (VR) catalogues, and therefore choose to re-find the centres of the haloes. To do this we first take the position as estimated by VR and then we use shrinking spheres to find the centre of mass of both the BCG (assuming that the stars are the dominant mass baryonic component within the central region) and the dark matter. We begin at $50$kpc and incrementally shrink to $1$kpc (or when we run out of particles). We then measure the offset between the two estimators and take the median of the distribution (as has been done previously) for a given dark matter model and redshift. Fig \ref{fig:wobble_z} shows the result with each panel showing the median offset as a function of cross-section for a given redshift (shown in the inset). To each figure we fit a linear function (in $\log(\sigma_{\rm DM}/m)$), parametrised by, 
\be
\mu = a +b\log\left(\frac{max\{\sidm,\sigma_{\rm eff}\}}{1{\rm cm}^2{\rm /g}}\right),
\label{eqn:wobble}
\ee
 where $a$ and $b$ are free parameters to fit and function inside the log takes the maximum value between the true cross-section and the effective cross-section due to coring by the artificial softening length of the simulation. \citet{SIDM_BAHAMAS} found that the effective cross-section for CDM was $\sigma_{\rm eff}\sim10^{-2}$ \cpg~, here the dark matter particle mass is almost two orders of magnitude smaller, resulting in a smoothing length $\sim5$ times smaller. We therefore choose an effective cross-section of $\sigma_{\rm eff}=2\times10^{-3}$ \cpg~. We find that altering this value does not substantially impact the fits in Figure \ref{fig:wobble_z}. We study the evolution of $a$- the intercept (or the cross-section at $\sigma_{\rm DM}/m=1.0$ \cpg) and $b$, the sensitivity of the wobble to the cross-section. The top panel of Figure \ref{fig:wobble_acc} shows the time evolution of the intercept as a function of look-back time, $t_{\rm lb}$ (and associated redshift on the top x-axis) and the middle panel shows the evolution of the wobble sensitivity. We find in both cases that they are very small at high redshift, with the sensitivity actually becoming negative. They both grow to peak around $z=0.25$ ($t_{\rm lb}=\sim3$ Gyr), however then the wobble decreases hereafter, counter to what we had expected with the growth of cores increasing over time (Figure \ref{fig:redshift}). \citet{darkgiants} had shown that what drives the wobbles are major mergers in cored halos, therefore we postulate that this wobble is in-fact driven by accretion of mass on to the halo. To understand this we derive the mass accretion rate, which is simply the change in halo mass of each cluster with respect to time, and a proxy for the core size, which is the ratio of the three-dimensional density at a radius of $10$kpc, $\rho_10$ and at $100$ kpc $\rho_100$. We show the mass accretion (solid lines) and the density ratio (dotted lines) of each simulation in the bottom panel. We find that the mass accretion peaks at $z\simeq0.5$, subsequently dropping off at lower redshifts. This is similar to other studies looking at merger rates of halos in simulations \citep{merger_rates}. We find that the shape of the mass accretion function is similar to that of both the wobble amplitude and the sensitivity, however with a small delay.  We hypothesise that this is because at $z=0.5$ the core has not sufficiently grown yet for the wobble to persist. The dotted line supports this showing how the core does not become significant until $z=0.375$.
To estimate the lag we fit a generalised model to the mean of the mass-accretion rate with a Gaussian Process Regressor \footnote{\url{https://scikit-learn.org}} along with a Rational Quadratic kernel that has best fitting meta-parameters of scale-length, $s=100$ and a $\alpha=6.3$. We show the corresponding fit as the dotted black line in the bottom panel of Figure \ref{fig:wobble_acc}. We hypothesise that it is this mass-accretion that is driving the relationship seen in the top two panels. We therefore adopt this shape and simply scale it and stretch it such that we fit 
\be
y = g(t_{\rm LB} - l)\cdot d+e,
\label{eqn:refit}
\ee
where, $g(x)$ is the normalised output of the Gaussian Process fitted to the mass accretion, $t_{\rm LB}$ is the look-back time, $l$ is the lag, and $d$ and $e$ are values to be fitted. We fit to both the wobble amplitude and the sensitivity and show the resulting fit in the sold red and its $1-\sigma$ error in the shaded region. We find in the case of the intercept (top panel of Figure \ref{fig:wobble_acc}) $l=1.7\pm0.7$ Gyr,$~d=2.4\pm0.6, ~e=2.46\pm0.34$, and for the middle panel, $l=2.6\pm0.6$ Gyr, $d=1.4\pm0.2, e=-0.5\pm0.02$, and in both cases finding a remarkably good fit. The rapid drop off in amplitude of the signal, suggests that the timescale for dynamical relaxation is much smaller growth rate of the wobble driven by mass accretion.

We also look at the offsets in CDM (shown as clear circles in the top panel of Figure \ref{fig:wobble_acc}) and find that offsets are not sensitive to mass accretion in the same way as SIDM for $z<0.75$ and trace closely the softening length of the simulation. Moreover, interestingly we find that the mass accretion in SIDM is lower at higher redshifts leading to a negative dependence of the Wobble on the cross-section, with lower redshift the wobble tracing the softening length of the simulation (black dashed line). This seems to be a combination of the decrease mass accretion and that CDM experiences lower density ratio at higher redshifts.

Given the coarse temporal resolution we have due to physical limitations of storage space of the snapshots we choose to carry out one simulation with extremely high fidelity snapshots for two models (CDM and SIDM0.2) of dark matter to observe the connection between mass accretion and the dark matter wobble. We show the results in Figure \ref{fig:hig_fid}. The first figure shows the mass accretion for the two models, showing that there is an order 1\% difference between them. It also shows how there is a large amount of structure accreted at $z\sim1.4$, as well at a later time at $z\sim0.2$. The second Figure shows the evolution of the offset between the dark matter and the BCG. We find that the offset for both models is consistent with each other until $z<0.5$,  when the SIDM0.2 model begins to exhibit oscillations at regular intervals with a peak amplitude of $\sim10$kpc. To better show this, we overlay the median filter in a dotted line. The third figure shows the ratio between the three dimensional density at $10$ kpc and $100$ kpc, $\Gamma=\rho_{10}/\rho_{100}$. We see a similar trend to Figure \ref{fig:wobble_acc} where at high redshifts SIDM0.2 seems to have a higher ratio, but as the halo evolves we clearly see the creation of the core at $z<0.5$. Finally we cross-correlate the median filtered offset with the mass accretion rate and show the result correlation, $\xi=\langle \dot{M}\cdot X\rangle$ as a function of the lag between the signals. We find that CDM exhibits no significant correlation between the two and  the peak correlation of SIDM0.2 lies at $t_{\rm lag}=1.2$ Gyr, roughly consistent with that found for the overall distribution in Figure \ref{fig:wobble_acc}. Moreover, we find at early times that despite the large amount of accreted mass, neither dark matter model experiences any kind of oscillation. However, at late times we find that now a core has been created, the increase in mass accretion does instigate a wobbling. This supports the hypothesis that the wobble is driven by mass accretion in a cored halo.

Finally to confirm the physical origin of the wobble (and not just increased error in the centring of the halo), we look at the velocity offset between the main halo and the stellar halo. Theoretically, should the BCG indeed wobble in the potential of the dark matter halo, the velocity should anti-correlate with the offset for SIDM and not for CDM. We explore this hypothesis and measure the relative velocity of the stars with respect to the centre of mass of the main halo (CoM). We then average this over all halos and all snapshots that have positive wobble ($z<0.75$). In order to account for differing halo mass (since the acceleration is dependent on the mass of the main halo and not the stellar halo), we rescale the velocities by the inverse-square-root, i.e $\langle v\rangle= \sqrt{M/10^{14}}$. Figure \ref{fig:wobble_vel} shows the results for three of our dark matter models. We see clearly here that the anti-correlation is strongest with the highest cross-section and that CDM is consistent with no correlation. We fit a linear function to this relationship and show the gradient of this as an inset. We find the significance of the anti-correlation for SIDM0.2 is 10 standard deviations from the null hypothesis that there is no correlation.  The intercept for the three plots are int$_{\rm CDM}=363\pm21$km/s,  int$_{\rm SIDM0.1}=418\pm14$km/s,  int$_{\rm SIDM0.2}=468\pm16$km/s

We note however, that the velocities of the dark matter halo is not directly observable. It is possible that the Intra Cluster Light (ICL) could trace this potential and therefore velocity estimates of the ICL could provide a way to trace its velocities. This is something for future work to look in to.

\section{Conclusions}\label{sec:conclusions}
We present a suite of one hundred zoom-in simulations of massive ($M/M_{\rm, 200} > 5\times10^{14}$) galaxy clusters simulated with five different self-interaction models including $\sidm=0.,0.01,0.05,0.1,0.2$ \cpg. We super-sampled our simulations reaching a dark matter / initial gas particle mass of $m_{\rm DM}/m_{\rm Gas}=\sim1/10$ (whereas normally it is $m_{\rm DM}/m_{\rm Gas}\sim5$). With this prescription and a dark matter particle mass of $0.68\times10^8M_\odot$ we are able to probe previously untested cross-sections. We calibrate our simulations on four key observables: the gas fraction within $R_{500}$, the stellar fraction within the same radius, the Brightest Cluster Galaxy (BCG) half mass radius versus its total stellar mass and the stellar mass to halo mass function of centrals. We are able to produce realistic galaxy clusters and as in \citet{RobertsonBAHAMAS} we find that there is no dependency of this calibration on the self-interaction cross-section, and hence does not require re-calibration. We present our results and find that the predicted three-dimensional density profiles predictably are less dense in the core of the cluster and tend to the same as collisionless dark matter in the outskirts. We find that CDM matches extremely well and we match the predictions from the BAHAMAS SIDM simulations down to the convergence radius of the simulations. We also find that even for the smallest cross-sections there is a discernible difference between the stellar and dark matter density profiles. We analyse the redshift dependence of the total matter density profile and find that the SIDM features only begin to reveal themselves in clusters at $z<0.5$, thus any survey attempting to constrain SIDM should limit to low redshift objects.

Following this we carry out an in-depth analysis of the cosmological ``wobbling'' signal, whereby the BCG oscillates in the bottom of an SIDM cored dark matter potential. We find that counter to expectations the three-dimensional wobbling signal peaks at $z=0.25$, dropping thereafter, despite the core-size continuing to increase. We measure the mass accretion rate of the halos in each cosmology and the core creation time and find that the wobble traces this rate remarkably well, albeit with a delay due to the late time creation of the dark matter core (that is required for the BCG to wobble).  Interestingly we find that SIDM0.2 has a delayed mass accretion. We find that the wobble amplitude falls off extremely quick at late times when mass accretion also drops, suggesting that these oscillations have relatively short relaxation times. We measure the lag between the mass accretion and the formation of a wobble to be $t_{\rm lag} = 2.6\pm 0.6$~Gyr. We follow this up by measuring the correlation between the separation of the BCG and the dark matter halo and the relative velocity of the two halos - a hallmark of harmonic oscillation. We find that SIDM0.2 exhibits an extremely strong anti-correlation, whereas CDM seems to have no such anti-correlation, once again supporting the existence of a physical oscillation of the BCG in SIDM cosmologies. This is encouraging as any observed offset between dark matter and the BCG (in any halo) is likely a smoking gun for SIDM.

With the expected arrival of all-sky surveys such as Euclid in the coming year, we will require robust predictions for possible dark matter models. Self-interacting dark matter is theoretically robust and empirically possible, and as such represents an ideal candidate for dark matter. However, if we are to exploit the forthcoming data we require theoretical predictions to match. This suite of simulations represents a pathway to delivering high-resolution particle dark matter simulations at a fraction of the CPU cost. By super-sampling the dark sector we can resolve faint scatterings and make robust predictions. Moreover, by simulating many more points in the dark matter parameter space we provide further data to train, test and validate our models. In order to facilitate the usefulness of these simulations we make them freely available and public to the community.

\section{Data availability}
Postage stamps of all models are available at \url{https://tinyurl.com/36nzxeej}. These data consist of two dimensional maps, binned to $20$kpc in resolution out to $2$Mpc (resulting in image dimensions of 100x100). We provide total mass maps, dark matter, x-ray emission maps and stellar distributions, resulting in 4 channels. These are projected in three axes resulting in a final shape of ($300\times4\times100\times100$). We provide snapshots 10, 9, 8 and 7, ($z=0.000$, $z=0.125$, $z=0.250$, and $z=0.375$) and for all SIDM models. For full particle data and catalogues please contact corresponding author.

\begin{acknowledgements}
This work was supported by the Swiss State Secretariat for Education, Research and Innovation (SERI) under contract number 521107294. 
This work was supported by a grant from the Swiss National Supercomputing Centre (CSCS) under project ID s1235 on Alps. The authors would like to thank Camila Correa without whom this project would not have got off the ground and John Biddiscombe for providing valuable insights and aid with the HPC at CSCS.
\end{acknowledgements}

\bibliographystyle{aa}
\bibliography{bibliography}
\label{lastpage}

%%%%%%%%%%%%%%%%%%%%%%
\begin{appendix}
\section{Further convergence tests}

\begin{figure}

    \begin{center} 

    \includegraphics[width=0.89\columnwidth]{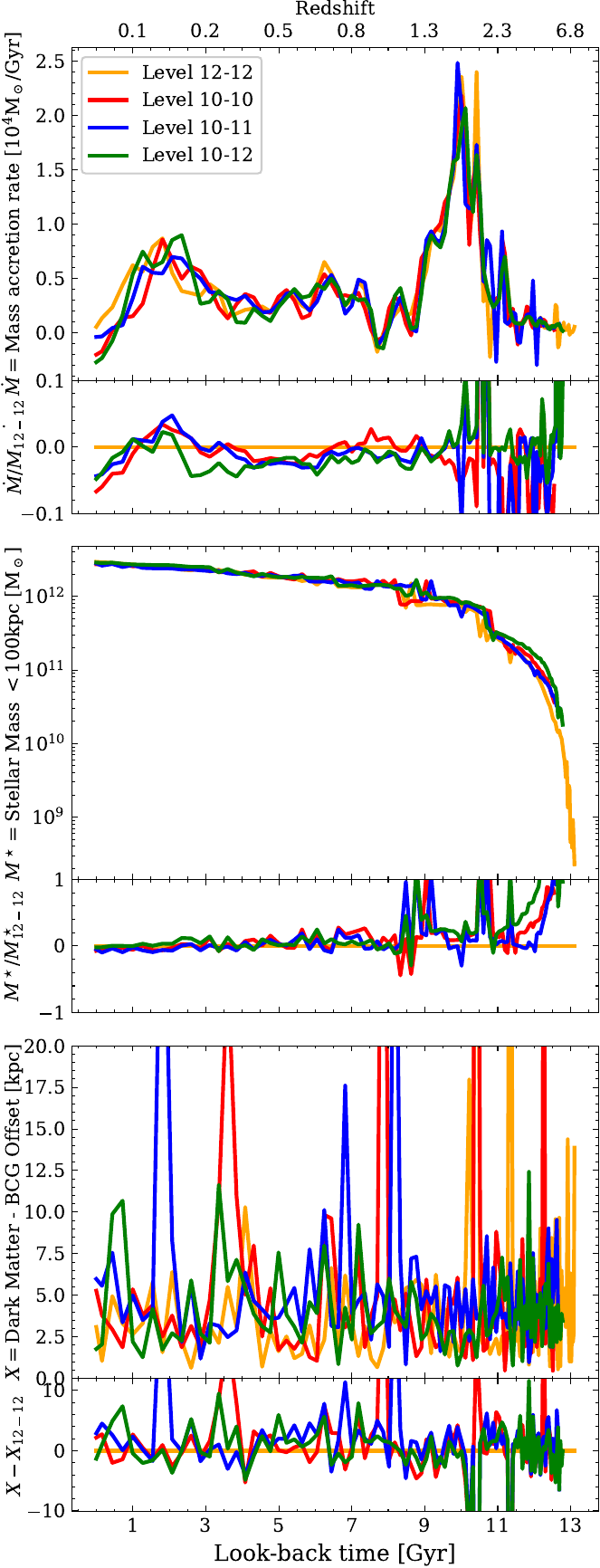}
    \caption{ Additional convergence tests: Each panel shows results from the same halo for four different resolutions as found in Figure \ref{fig:super_sampled}, where the first number corresponds to the level used in {\tt MUSIC} for the gas mass and the second for the dark matter. Each figure shows the absolute value in the top panel and the value relative to the high resolution, level 12-12, simulation in the bottom panel. From left to right we show the mass accretion rate, the stellar mass within $100$kpc, and the dark matter - brightest cluster galaxy offset. As expected we find in all cases that there is no clear difference between the different runs.}
    \label{fig:further_conv}
    \end{center}
    \end{figure}

\begin{figure}
    \begin{center}
    \includegraphics[width=\columnwidth]{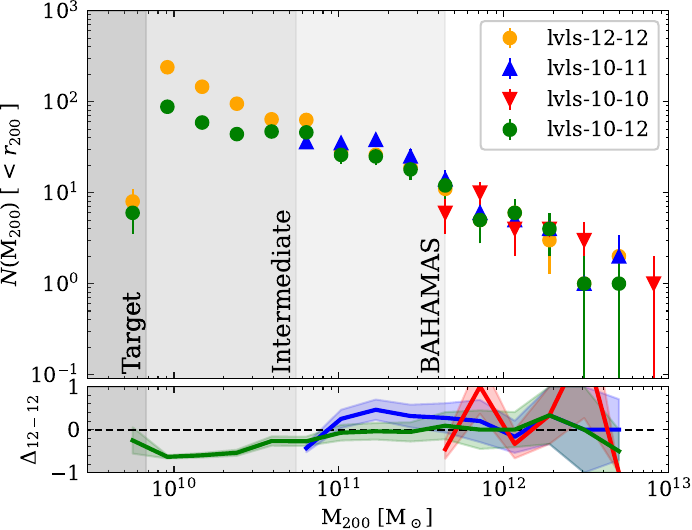}
    \caption{ Consistency of the sub-halo mass function. We count the number of mass halos inside one virial radii of the cluster and show their mass distribution here. Assuming a minimum particle number of 100 particles for a halo we show the mass resolution limit of each simulation where BAHAMAS is roughly equal to levels 10-10, intermediate corresponds to levels 10-11 and target is levels 10-12. The bottom panel shows the ratio between the mass function and the highest resolution, non-super-sampled run (12-12). We find that the sub-halo mass functions for the target resolution is moderately suppressed below $M_{200}<2\times10^{10}M_\odot$.}
    \label{fig:subhalo_conv}
    \end{center}
    \end{figure}

We carry out further convergence tests to ensure that our simulations do not suffer from any numerical artefacts. We simulate the same halo four times with the same resolution levels as in Figure \ref{fig:super_sampled}. We label each simulation according to their {\tt MUSIC} initial conditions level, with the first number referring to the baryonic / initial gas mass resolution and the second the dark matter particle resolution. For these convergence tests we output many more snapshots in order to get a high-fidelity view of the growth of our halos. We analyse four key properties from the output including: mass accretion rate (left hand panels of figure \ref{fig:further_conv}), the stellar mass growth of the central halo (i.e. the star formation rate) in the middle panels of figure \ref{fig:further_conv}, and the dark matter - brightest cluster offset for each resolution over the entire cosmic time. The final test is the consistency of sub-halo mass function. Figure \ref{fig:subhalo_conv} shows the mass function for sub-halos inside one virial radii of the cluster.  We find that the sub-halo mass functions for the target resolution is moderately suppressed below $M_{200}<2\times10^{10}M_\odot$. It is difficult to interpret whether this is due to the super-sampling, or due to the halo-finder struggling to identify halos with significantly different particle masses. We found that chosen parameters for the {\tt VELOciraptor} halo finder resulted in a broad range of predicted mass functions for the target resolution. Despite the moderate suppression, halos of this mass will not affect the results of this work which focuses on the main halo properties. Future work studying sub-halos in super-sampled SIDM should consider looking in to this further.

We do notice that the dark matter - brightest cluster galaxy offsets can sometimes be very noisy with offsets greater than $20$ kpc observed. These outliers occur either in simulations of low resolution where halos are hard to separate, of at times of very high mass accretion at early times when again halo separation is difficult. Indeed, our target resolution does not suffer from these outliers. We therefore conclude that our simulations do not appear to suffer from any artefacts due to super-sampling of the dark matter.

 \FloatBarrier    
\section{Contamination from boundary particles}

\begin{figure}
    \begin{center}
    \includegraphics[width=\columnwidth]{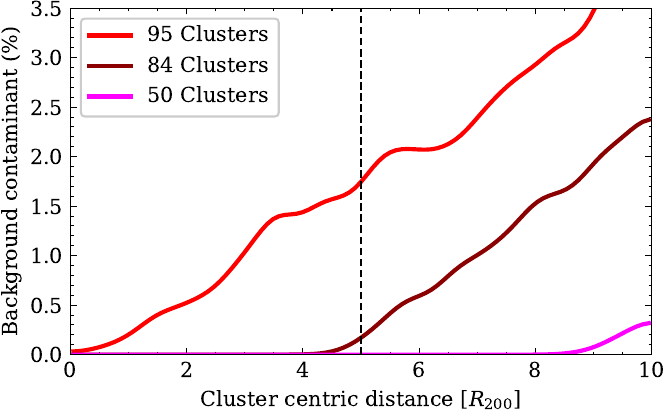}
    \caption{ Boundary particles, or low resolution particles that trace the large-scale structure should not enter in to the high resolution region. However, in rare cases this can occur. We show here the level of contamination \eqref{eqn:boundary}, in the main halo of boundary particles as a function of radius from the centre of the halo. Each line shows the maximum contamination for a given number of halos. So for example, half of the halos have negligible contamination, (orange line) with 95\% of our clusters having less than 2\% contamination at 5 virial radii (dark red line). }
    \label{fig:boundary}
    \end{center}
    \end{figure}

Boundary particles exist in the simulation to trace the large scale tidal fields that impact the growth of structure in the high-resolution regime. These particles theoretically should never actually enter this regime. However, when the initial Lagrangian region from which the initial conditions is cut, is too small, contamination can occur. In order to understand the contamination in our high-resolution we identify the boundary particles within the high-resolution region at z=0. We estimate the contamination by boundary particles and in Figure \ref{fig:boundary} show this a function of cluster centric radii (expecting increase in contamination as you tend to the boundary edge), where we define the contamination, $c$ as:
\be
c    = \frac{ n_{\rm b}}{n_{\rm b} + n_{\rm d}},
\label{eqn:boundary}
\ee
where $n_{\rm b}$ and $n_{\rm d}$ are the number of boundary and dark matter particles respectively. 

\FloatBarrier
\section{ Effective velocity-dependent models}

Simulating velocity-{\it independent} models, is qualitatively the same at simulating velocity-{\it dependent} models, for a given halo mass and redshift. We show here the effective models we simulate here ( for a given halo mass at a $z=0$ ), assuming that relative particle velocities follow $v\sim\sqrt{GM/r}$. The stars show the velocity-{\it dependent} models, and the lines show how these can be recast in to velocity-{\it dependent}, assuming a constant asymptotic cross-section (top panel) or a constant turn over velocity  (bottom panel). The caveat here is that the sub-halos remain unchanged in these models, whereas in a true velocity-{\it dependent} model, the lower particle velocities in these halos would change their mass distribution slightly.

\begin{figure}
\begin{center}
\includegraphics[width=0.5\textwidth]{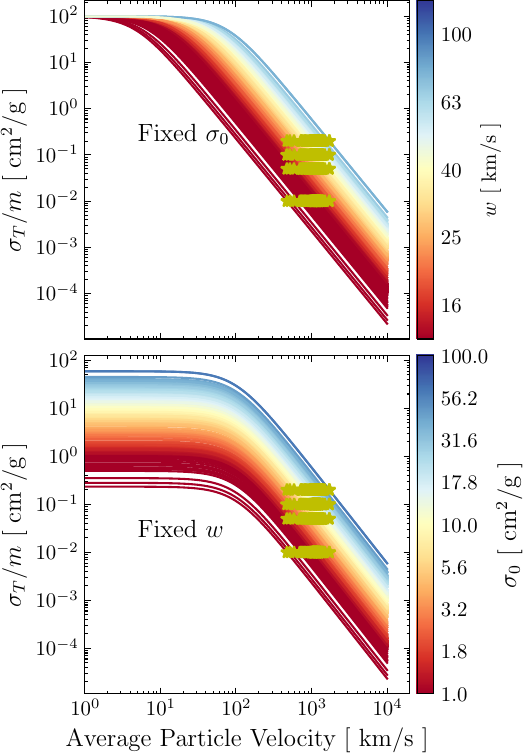}
\caption{ The effective models we simulate here ( for a given halo mass at a $z=0$ ), assuming that relative particle velocities follow $v\sim\sqrt{GM/r}$. The stars show the velocity-{\it independent} models, and the lines show how these can be recast in to velocity-{\it dependent}, assuming a constant asymptotic cross-section (top panel) or a constant turn over velocity (bottom panel). }
\label{fig:par_space}
\end{center}
\end{figure} 

\end{appendix}
\end{document}